\documentclass[submission, Phys]{SciPost}

\usepackage[caption=false,margin=0pt]{subfig}
\usepackage{graphicx}
\usepackage{amsmath,amsthm,amssymb}
\usepackage{bm,bbm,mathbbol}
\renewcommand\Im{\operatorname{Im}}
\renewcommand\Re{\operatorname{Re}}

\newcommand{\eq}[1]{Eq.~(\ref{eq:#1})}
\newcommand{\eqs}[2]{Eqs.~(\ref{eq:#1}) and~(\ref{eq:#2})}

\newcommand{\fig}[1]{Fig.~\ref{fig:#1}}
\newcommand{\sect}[1]{Sec.~\ref{sec:#1}}

\def\beq{\begin{equation}}
\def\eeq{\end{equation}}
\def\bea{\begin{eqnarray}}
\def\eea{\end{eqnarray}}

\def\ket#1{\vert#1\rangle}

\def\ip#1#2{\langle#1\vert#2\rangle}
\def\me#1#2#3{\langle#1\vert#2\vert#3\rangle}

\def\E{{\mathcal E}}

\def\ww{\omega}

\def\dk{[d\kk]}

\def\kk{{\bm k}}
\def\pp{{\bm p}}
\def\rr{{\bm r}}
\def\dd{{\bm d}}
\def\Aa{{\bm A}}
\def\RR{{\bm R}}
\def\OO{{\bm 0}}

\def\btau{{\bm\tau}}
\def\W{^{\rm (W)}}

\def\nn{\nonumber\\}

\def\AAAbold{\boldsymbol{\mathbbm{A}}}

\def\aaabold{\boldsymbol{\mathbbm{a}}}

\def\bnabla{{\boldsymbol\nabla}}

\def\la{\langle\kern-2.5pt\langle}
\def\ra{\rangle\kern-2.5pt\rangle}
\def\vt{\vert\kern-1.5pt\vert}
\usepackage{soul}  

\newcounter{comm} 

\begin{document}

\begin{center}{\Large \textbf{Assessing the role of 
        interatomic position matrix elements
    in tight-binding calculations of optical
    properties }}
\end{center}

\begin{center}
Julen Iba\~{n}ez-Azpiroz\textsuperscript{1,2*},
Fernando de Juan\textsuperscript{2,3},
Ivo Souza\textsuperscript{1,2}
\end{center}

\author{Julen Iba\~{n}ez-Azpiroz} 

\author{Fernando de Juan}

\author{Ivo Souza} 

\begin{center}
{\bf 1} Centro de F{\'i}sica de Materiales, Universidad
  del Pa{\'i}s Vasco, 20018 Donostia-San Sebasti{\'a}n, Spain
\\
{\bf 2} Ikerbasque Foundation, 48013 Bilbao, Spain
\\
{\bf 3} Donostia International Physics Center, 20018 
 Donostia-San Sebasti\'{a}n, Spain
\\
* julen.ibanez@ehu.es
\end{center}

\date{\today}

\section*{Abstract}

{\bf We study the role of
   hopping matrix elements of the position operator $\hat\rr$ in
  tight-binding calculations of linear and nonlinear optical
  properties of solids.  Our analysis relies on a
  Wannier-interpolation scheme based on \textit{ab initio}
  calculations, which automatically includes matrix elements of
  $\hat\rr$ between different Wannier orbitals. A common
  approximation, both in empirical tight-binding and in
  Wannier-interpolation calculations, is to discard those matrix
  elements, in which case the optical response only depends on the
  on-site energies, Hamiltonian hoppings, and orbital centers.  We
  find that interatomic $\hat\rr$-hopping terms
   make a sizeable contribution to the shift
  photocurrent in monolayer BC$_2$N, a covalent acentric crystal.  If a
  minimal basis of $p_z$ orbitals on the carbon atoms is used to model
  the band-edge response, even the dielectric function becomes
  strongly dependent on those terms.}

\vspace{10pt}
\noindent\rule{\textwidth}{1pt}
\tableofcontents\thispagestyle{fancy}
\noindent\rule{\textwidth}{1pt}
\vspace{10pt}

\section{Introduction}
\label{sec:introduction}

Empirical tight-binding (TB) is the method of choice for obtaining a
simple and intuitive description of the electronic structure of
solids~\cite{harrison-book80}.  In this method, the basis is defined
implicitly through the on-site energies and the Hamiltonian
hopping matrix elements,
\beq
\label{eq:e-t}
\me{\OO n}{\hat H}{\RR m}=
{\epsilon_m\delta_{\OO\RR}}\delta_{nm}+t_{nm}(\RR)\,,
\eeq
without an explicit real-space representation of the basis
orbitals.\footnote{As our focus will be on orthogonal TB models with
  Wannier functions (WFs) as basis orbitals, we adopt the notation
  $\ket{\RR m}$ that is widely used in the literature to denote the
  $m$th Wannier orbital in the unit cell labeled by lattice vector
  $\RR$.}  While that information is already sufficient to evaluate
many physical quantities (energy bands, elastic constants, phonon
spectra, etc.), the calculation of
optical responses requires, in addition, the matrix elements of the
position (dipole) operator $\hat\rr$ in the TB basis.

In TB calculations of optical properties, it is customary to make the
simplest possible approximation for the position matrix; namely,
to discard all matrix elements of
  $\hat\rr$ between different Wannier orbitals, which we will refer to
  as the ``hopping matrix elements of $\hat\rr$'', or simply
  ``$\hat\rr$
  hoppings.''  When doing so, the only spatial information that is
  retained in the model are the orbital centers,
\beq
\label{eq:diag}
\me{\OO n}{\hat \rr}{\RR m}\approx
\btau_m\delta_{\OO\RR}\delta_{nm}\,,
\eeq
where $\btau_m$ is the center of the $m$th Wannier orbital in the home
cell.  This minimal spatial embedding of a TB model already allows to
incorporate electromagnetic fields in a gauge-invariant
manner~\cite{graf-prb95}. It is, nonetheless, a rather uncontrolled
approximation: symmetry-allowed intra-atomic matrix elements such as
$\me{s}{\hat{x}}{p_{x}}$ are discarded along with interatomic matrix
elements, all of which can in principle contribute to the optical
response.

The above approximation has been discussed in the
literature~\cite{bennetto-prb96,pedersen-prb01,PhysRevB.66.165212},
and the general prescription for incorporating $\hat\rr$-hopping terms
into both orthogonal~\cite{wang-prb06} and
non-orthogonal~\cite{lee-prb18} TB models has been described.
However, the impact of those terms on calculations of optical
responses has not been thoroughly examined.  An important step was
taken in Refs.~\cite{pedersen-prb01,PhysRevB.72.125105}, which
examined the corrections from intra-atomic {$\hat\rr$ hoppings} to the
linear optical response of a toy model.

In this work, we revisit the problem from an {\it ab initio}
perspective.  We restore the hopping terms that are missing from
\eq{diag},
\beq
\label{eq:r-full}
\me{\OO n}{\hat \rr}{\RR m}=
\btau_m\delta_{\OO\RR}\delta_{nm}+\dd_{nm}(\RR)\,,
\eeq
and employ a first-principles-based WF method to assess their
contribution to the linear and
quadratic
optical responses of a single
layer of the graphitic material
BC$_2$N~\cite{LWC89,WIM95,PhysRevB.73.193304}.

Our calculations use the Wannier interpolation
method~\cite{wang-prb06} and they proceed as follows.  After
performing an \textit{ab initio} calculation of the electronic
structure, we construct, in a post-processing step, well-localized WFs
spanning the relevant bands.  We then take those WFs and use them as
an orthogonal TB basis to evaluate the band structure and the optical
matrix elements.  Since the Wannier orbitals are constructed
explicitly, the $\hat\rr$ hoppings can be tabulated and included,
along with the on-site energies, orbital centers, and $\hat H$
hoppings, in the calculation of optical responses; by selectively
discarding some or all of the $\hat\rr$ hoppings, we are able to gauge
their contributions.

The optical responses analyzed in this work are the dielectric
function and the shift photoconductivity.  The
latter is a quadratic response associated with a
shift in the center of mass of an electron as it is optically excited
from a valence band to a conduction band in a piezoelectric
crystal~\cite{Baltz1981,belinicher-jetp82,sipe-prb00}, and is known to
be quite sensitive to the spatial embedding of the TB
model~\cite{presting-pssb82,cook-nc17}. (The same happens with the
ground-state electric polarization~\cite{bennetto-prb96}, which is
associated with the center of mass of valence
electrons~\cite{king-smith-prb93}.)

Our test system, a single layer of BC$_2$N, was chosen for the
following reasons. (i) Its structure is noncentrosymmetric, a
necessary condition for observing a quadratic response.  (ii) As it is
a covalent crystal with strong orbital overlap between different
sites, interatomic terms can be expected to play a significant
role. (iii) Its simple band structure near the fundamental gap allows
for a complementary study based on a two-band $\kk\cdot\pp$ model
constructed from the Wannier Hamiltonian, which further highlights the
impact of approximation~\eqref{eq:diag}.

The paper is organized as follows.  In \sect{responses} we review the
expressions for the dielectric function and shift photoconductivity in
the independent-particle approximation.  The connection between
Wannier interpolation and TB theory, and the contribution of $\hat\rr$
hoppings to the optical matrix elements, is discussed in
\sect{interpol}.  In \sect{comp-details} we describe the \textit{ab
  initio} and Wannier-interpolation calculations, and in
\sect{results} we present and analyze the numerical results.  We
conclude in \sect{discussion} with a summary and discussion, and
in Appendix~\ref{appendix} we describe how the $\kk\cdot\pp$ model is
constructed.

  \section{Dielectric function and shift photoconductivity}
\label{sec:responses}

In a nonmagnetic material such as BC$_2$N, the dielectric
  function is a symmetric tensor whose imaginary part is absorptive,
  with an interband contribution given by~\cite{sipe-prb00}
\beq
\label{eq:dielectric}
\epsilon^{\prime\prime}_{ab}(\ww)=\frac{\pi e^2}{\hbar}
\int\dk\sum_{nm}f_{\kk nm}\,{\Re}\left(r^a_{\kk nm}r^b_{\kk mn}\right)
\delta(\omega_{\kk mn}-\omega)\,.
\eeq
Here $e>0$ is the elementary charge, $f_{\kk nm}=f_{\kk n}-f_{\kk m}$
and $\hbar\omega_{\kk nm}=E_{\kk n}-E_{\kk m}$ 
are differences between
occupation factors and between band energies, $\dk=d^dk/(2\pi)^d$ in
$d$ dimensions, and the integral is over the first Brillouin zone
(BZ).  Finally,
\beq
\label{eq:r}
\rr_{\kk nm}=(1-\delta_{nm})\Aa_{\kk nm}
\eeq
is the interband dipole given by the off-diagonal part of the Berry
connection matrix
\beq
\label{eq:A}
\Aa_{\kk nm}=i\ip{u_{\kk n}}{\bnabla_\kk u_{\kk m}}\,,
\eeq
where $\ket{u_{\kk m}}$ denotes the cell-periodic part of a Bloch
state $\ket{\psi_{\kk m}}$.

Noncentrosymmetric crystals display a nonlinear optical effect
  known as the bulk photovoltaic (or photogalvanic) effect, which can
  be divided phenomenologically into ``linear'' and ``circular''
  parts~\cite{belinicher_photogalvanic_1980,sturman-book92,ivchenko-book97}.
  The linear bulk photovoltaic effect occurs in piezoelectric crystals
  such as BC$_2$N, and is described by the relation
\beq
\label{eq:lbpve}
j_a=2\sigma_{abc}(0;\ww,-\ww)\E_b(\ww)\E_c(-\ww)\,.
\eeq
The shift current corresponds to the interband part of this response
and is given by~\cite{sipe-prb00}
\bea
\label{eq:sigma-abc}
\sigma_{{abc}}(0;\ww,-\ww)=\dfrac{\pi e^3}{{2}\hbar^2}
\int\dk\sum_{nm}f_{\kk nm}\,
{\Im}\left[r^b_{\kk mn}r^{c;a}_{\kk nm}{+(b\leftrightarrow c)}\right]
\delta(\omega_{\kk mn}-\omega)\,,
\eea
where
\beq
\label{eq:gen-der}
r^{c;a}_{\kk nm}=\partial_a r^c_{\kk nm}
-i\left(A^a_{\kk nn}-A^a_{\kk mm}\right)r^c_{\kk nm}
\eeq
denotes the {gauge-covariant} ``generalized derivative'' of the
interband dipole {with respect to $\kk$ ($\partial_a=\nabla_{k_a}$
  is the ordinary $\kk$ derivative)}.

\section{Wannier interpolation}
\label{sec:interpol}

\subsection{Energy bands}
\label{sec:bands}

After the {\it ab initio} total-energy calculation, we construct in a
post-processing step a set of well-localized WFs.  These are chosen to
span a group of bands that includes the initial and final states
involved in interband absorption processes up to some desired
frequency.  From the Wannier orbitals we then define a set of Bloch
basis states as
\beq
\label{eq:u-w}
\ket{\psi\W_{\kk m}}=
\sum_\RR\,e^{i\kk\cdot(\RR+\btau_m)}\ket{\RR m}\,,
\eeq
take matrix elements of the \textit{ab initio} Hamiltonian between
those states,
\bea
\label{eq:H-w}
H\W_{\kk nm}=
\me{\psi\W_{\kk n}}{\hat H}{\psi\W_{\kk m}}
=\sum_\RR\,e^{i\kk\cdot(\RR+\btau_m-\btau_n)}
\me{\OO n}{\hat H}{\RR m}\,,
\eea
and diagonalize the resulting matrix,
\beq
\label{eq:eig}
\left(U^\dagger_\kk H\W_\kk U_\kk\right)_{nm}=E_{{\kk m}}\delta_{nm}\,.
\eeq
With a proper choice of WFs, the eigenvalues $\{ E_{{\kk m}}\}$
provide a smooth interpolation across the BZ of the selected group of
density functional theory (DFT)
bands~\cite{souza-prb01}.

\subsection{Optical  matrix elements}

The same interpolation strategy can be applied to other $k$-space
quantities. In particular, the
transformed cell-periodic states
\beq
\label{eq:u-H}
\ket{u_{\kk n}}=\sum_m\,\ket{u\W_{\kk m}}U_{\kk mn}
\eeq
interpolate the {\it ab initio} cell-periodic eigenstates, allowing to
treat wavefunction-derived quantities such as the Berry
connection~\cite{wang-prb06}.  Inserting \eq{u-H} in \eq{A} yields
\begin{subequations}
\label{eq:A-wannier}
\begin{align}
\label{eq:A-decomp}
\Aa_\kk&=\AAAbold_\kk+\aaabold_\kk\,,\\
\label{eq:AAA}
\AAAbold_\kk&=i U^\dagger_\kk\bnabla_\kk U_\kk\,,\\
\label{eq:aaa}
\aaabold_\kk&=U^\dagger_{\kk}\Aa\W_\kk U_\kk\,,
\end{align}
\end{subequations}
where
\bea
\label{eq:A-w}
\Aa\W_{\kk nm}=
i\ip{u\W_{\kk n}}{\bnabla_\kk u\W_{\kk m}}
&=&\sum_\RR\,e^{i\kk\cdot(\RR+\btau_m-\btau_n)}
\me{\OO n}{\hat \rr-\btau_m}{\RR m}\nn
&=&{\sum_\RR\,e^{i\kk\cdot(\RR+\btau_m-\btau_n)}\dd_{nm}(\RR)}\,,
\eea
and \eq{r-full} was used in the last step.  We will refer to
$\AAAbold_\kk$ and $\aaabold_\kk$ as the ``internal'' and ``external''
parts of the Berry connection matrix $\Aa_\kk$.\footnote{The
  generalized derivative of the {interband dipole matrix
    [\eq{gen-der}]} also splits into internal and external parts when
  expressed in the Wannier representation, see
  Ref.~\cite{azpiroz-prb18} for details.}

The convention adopted in \eq{u-w}, with the phase factor
$e^{i\kk\cdot\btau_m}$ included in the Bloch sum, is the most natural
one for dealing with geometric quantities in $k$
space~\cite{vanderbilt-book18}. With that convention, when all
$\hat\rr$ hoppings $\dd_{nm}(\RR)$ are discarded \eq{A-w} for
$\Aa\W_\kk$ vanishes identically
so that $\Aa_\kk$ reduces to the
  $\AAAbold_\kk$ term given by~\cite{wang-prb06}
\beq
\label{eq:A-int}
\AAAbold_{\kk nm}=
\frac{
  \left[
  U^\dagger_\kk \left(\frac{1}{\hbar}\bnabla_\kk H\W_\kk\right)U_\kk
  \right]_{nm}}
{i\omega_{\kk nm}}\qquad (m\not= n)\,,
\eeq
where the numerator is an ``effective'' TB velocity matrix
  element obtained by substituting
  $\hat{\bm v}\rightarrow (1/\hbar) \bnabla_\kk
  H\W_\kk$~\cite{graf-prb95}. The $\AAAbold_\kk$ term is ``internal''
  in that it only depends on the minimal TB ingredients contained in
  \eqs{e-t}{diag}: on-site energies, (interatomic)~$\hat H$ hoppings,
  and Wannier centers.

Let us now restore {the ``external'' $\hat\rr$-hopping terms}, and
examine their contributions to $\Aa_\kk$. Intra-atomic $\hat\rr$
hoppings make a $\kk$-independent contribution to \eq{A-w}, while the
contribution from the interatomic ones varies with $\kk$.\footnote{The
  distinction between intra- and inter-atomic hoppings can only be
  made when the Wannier orbitals are centered on the atoms,
  as is the case in the present study.}  In the limit where the
overlap between orbitals on different sites is negligibly small, all
interatomic terms vanish and \eq{A-wannier} reduces to
$\Aa_\kk\approx U^\dagger_\kk \dd^{({\rm intra})} U_\kk$, where
$\dd^{({\rm intra})}$ denotes the intra-atomic part of the
$\hat\rr$-hopping matrix $\dd(\RR)$. Away from that limit, interatomic
$\hat H$- and $\hat\rr$-hopping terms give competing contributions to
$\Aa_\kk$.

Thus, the standard treatment of optical properties in TB can be
systematically improved by progressively adding
$\hat\rr$-hopping terms to the model in a sequence of steps:
\begin{enumerate}
\item Only include {on-site energies, orbital centers, and
  $\hat H$ hoppings}.
\item Add  intra-atomic {$\hat\rr$ hoppings} (if any).
\item Add nearest-neighbor interatomic {$\hat\rr$ hoppings}.
\item Add next-nearest-neighbor {$\hat\rr$ hoppings}.
\item  ...
\end{enumerate}
Step 1 corresponds to the uncorrected TB model, step 2 adds the
intra-atomic corrections, and steps 3 and higher incorporate
interatomic corrections. Intra-atomic corrections are expected to be
important for optical transitions involving localized $d$
orbitals~\cite{pedersen-prb01}. The role of interatomic
corrections remains less clear, as they were not considered in some of
the previous works; they will be included below in our study of
monolayer BC$_2$N.

\section{Structural and computational details} 
\label{sec:comp-details}

\begin{figure}[b!]\centering
\includegraphics[width=0.7\linewidth]{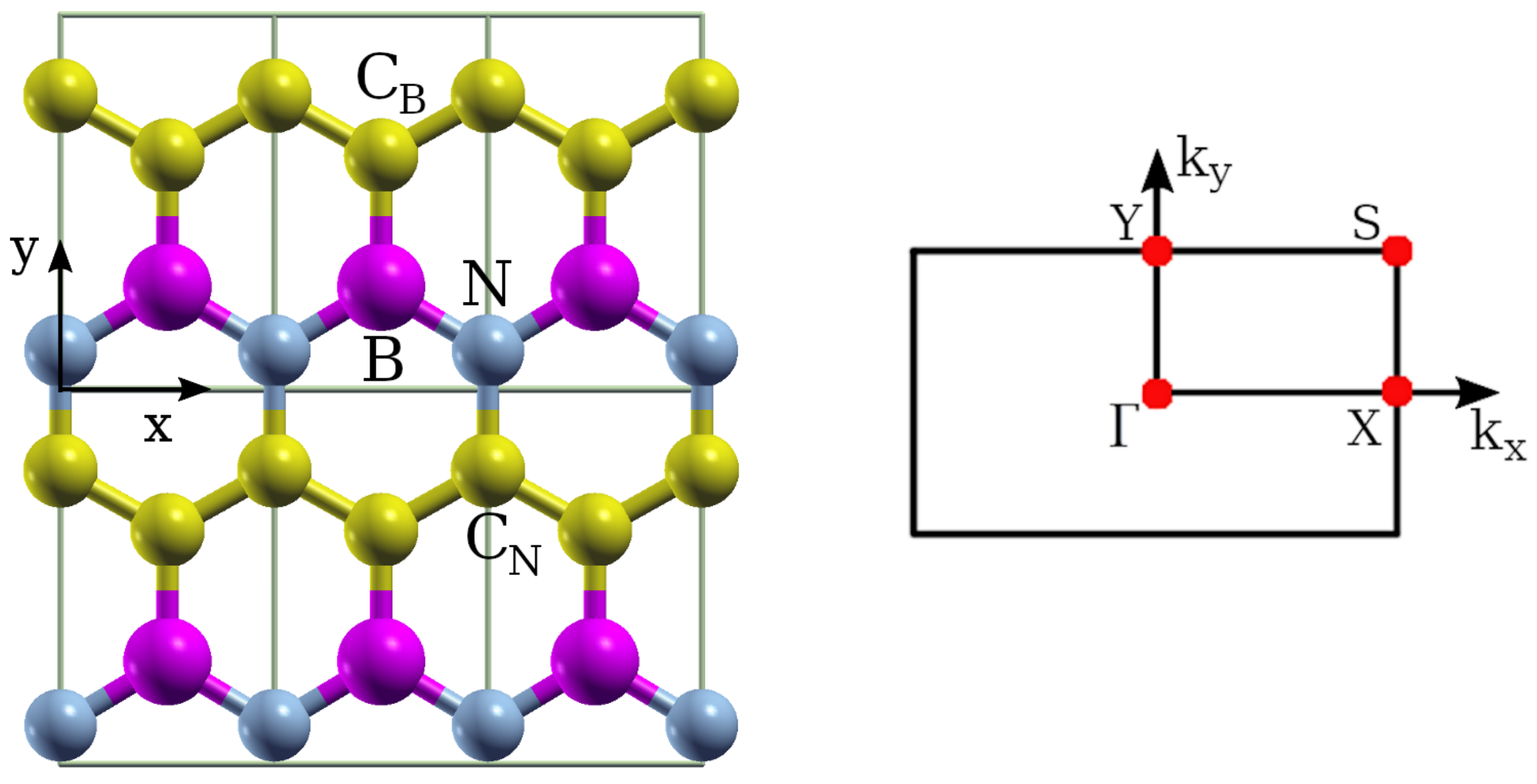}
\caption{Crystal structure and Brillouin zone of monolayer BC$_{2}$N.
  There is one formula unit per cell with two inequivalent carbon
  atoms, and we use the symbol C$_{\mathrm{N}}$ (C$_{\mathrm{B}}$) to
  label a carbon atom with a nitrogen (boron) atom among its nearest
  neighbors.}
\label{fig:bc2n-crystal-str}
\end{figure}

The structure of monolayer BC$_{2}$N, depicted in
\fig{bc2n-crystal-str}, is formed by alternating zigzag chains of
carbon and boron nitride {atoms}.  The unit cell contains four atoms;
the two carbon atoms are inequivalent, and we use the symbols
C$_{\mathrm{N}}$ and C$_{\mathrm{B}}$ to label the ones with a
nitrogen and a boron atom among their nearest neighbors (NNs),
respectively.  The structure is polar along $\hat{\textbf{y}}$, and
has mirror symmetries $M_{x}$ and $M_{z}$.  The BZ is also shown in
\fig{bc2n-crystal-str}, with the high-symmetry points indicated.

The DFT calculations are performed using
the {\tt Quantum ESPRESSO} code package~\cite{gianozzi-jpcm09}.  We
treat the core-valence interaction using scalar-relativistic projector
augmented-wave pseudopotentials taken from the {\tt Quantum ESPRESSO}
website.  The pseudopotentials are generated with the
Perdew-Burke-Ernzerhof exchange-correlation
functional~\cite{perdew-prl96}, and the energy cutoff for the
plane-wave basis expansion is set at 70~Ry.

To study the monolayer we employ a slab of length $l=$20~\AA\, along
the out-of-plane direction, and take the lattice constants
($a_{1}=2.46$~\AA~along $\hat{\bm x}$ and $a_{2}=4.32$~\AA~along
$\hat{\bm y}$) from Ref.~\cite{PhysRevB.73.193304}.  For the
self-consistent calculation we use a $10\times 10\times 1$
\textit{k}-point mesh, and switch to a $20\times 20\times 1$ mesh for
the non-self-consistent step from which we obtain the Bloch functions
that are used as input for the Wannierization
procedure~\cite{marzari-prb97,souza-prb01}.

\begin{figure}[t!]\centering
\includegraphics[width=0.9\linewidth]{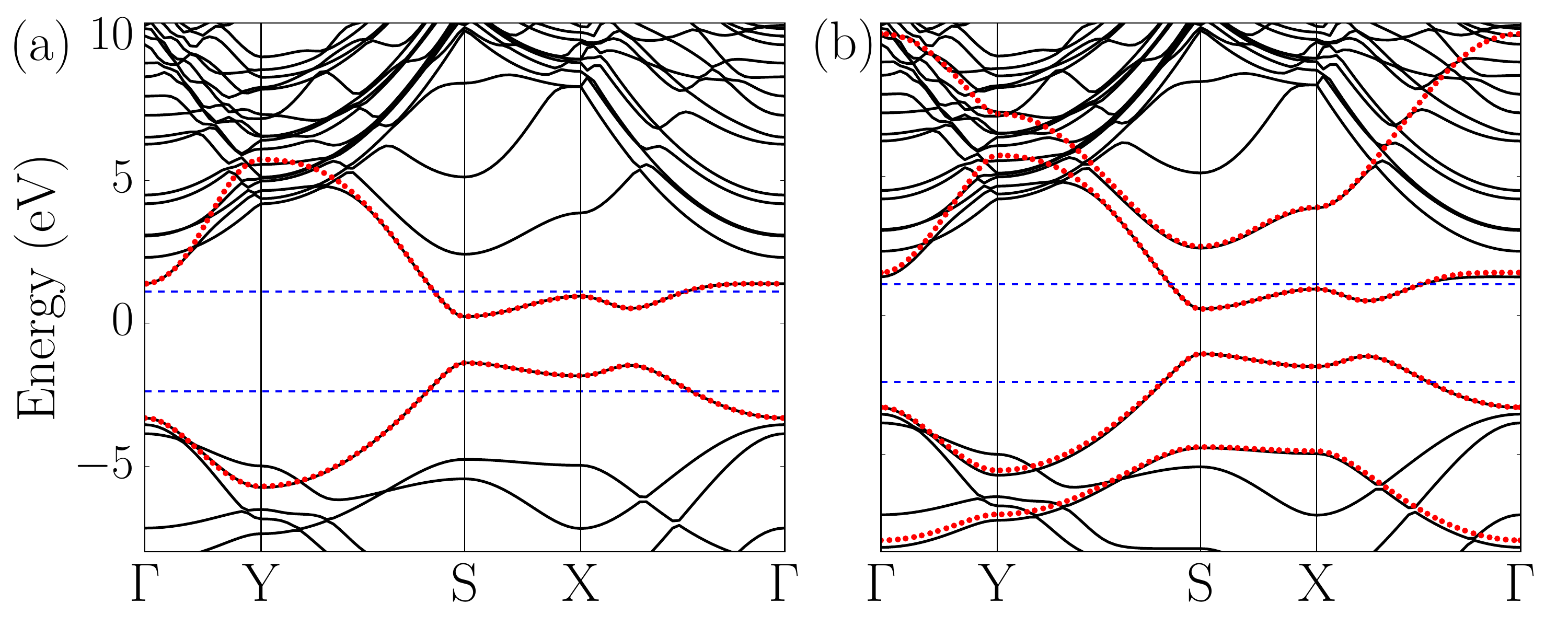}
\caption{(a) and (b) {show the} DFT bands (black lines) and
  Wannier-interpolated bands (red dots) for the {basis} set with two
  and four WFs {per cell}, respectively.  The dashed blue lines
  indicate the {boundaries of the} inner disentanglement {energy}
  window~\cite{souza-prb01}.  Energies are measured from the Fermi
  level.}
\label{fig:bands-comparison}
\end{figure}

We use the {\tt Wannier90} code package~\cite{pizzi_wannier90_2019} to
generate atom-centered WFs spanning the states around the Fermi
level. To that end, we employ a one-shot projection
scheme~\cite{marzari-prb97} combined with band
disentanglement~\cite{souza-prb01}. We consider two different sets of
WFs: one with a $p_z$ orbital on every atom (with an average quadratic
spread of 1.08 \AA$^{2}$), and another with $p_z$ orbitals on the
carbon atoms only (with an average spread of 3.39 \AA$^{2}$). The
corresponding Wannier-interpolated bands are plotted in the two panels
of \fig{bands-comparison} along with the DFT bands.

To obtain well-converged Wannier-interpolated spectra for the
dielectric function and shift photoconductivity, we employ a dense
$2000\times 2000\times 1$ \textit{k}-point interpolation grid.  We use
a fixed width of 0.01~eV when broadening the delta functions in
Eqs. (\ref{eq:dielectric}) and (\ref{eq:sigma-abc}), and apply a
broadening of 0.04 eV to regularize the contribution of intermediate
states to the shift current~\cite{azpiroz-prb18}.  The occupation
factors are evaluated at $T=0$~K.

\begin{figure}[t]\centering
\includegraphics[width=0.9\linewidth]{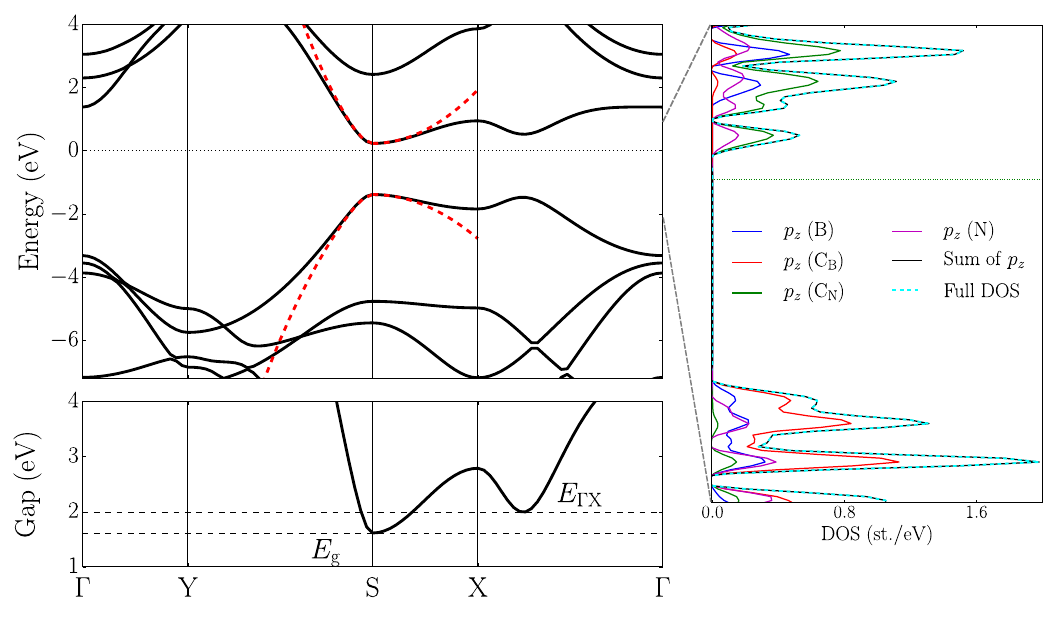}
\caption{Upper-left panel: solid black lines denote the
  DFT band dispersion near the Fermi
  level (taken as the zero of energy, dotted line).  Red dashed lines
  denote the dispersion {around the band edge} obtained from a
  two-band $\kk\cdot\pp$ model (see Sec.~\ref{sec:kp}).  Lower-left
  panel: direct energy gap between conduction and valence bands.
  $E_{\rm g}$ is the absolute minimum at S, and
  $E_{\Gamma \mathrm{X}}$ is a local minimum along the
  $\Gamma$--$\mathrm{X}$ line.  Right panel: density of states (DOS)
  around the Fermi level projected onto $p_z$ orbitals on each atom in
  the unit cell, compared with the full DOS.}
\label{fig:elec-str}
\end{figure}

\section{Numerical results}
\label{sec:results}

\subsection{Electronic structure and optical spectra}
\label{sec:elec-opt}

The DFT band structure of monolayer
BC$_2$N {(black lines in \fig{bands-comparison})} is replotted in
\fig{elec-str}
over a narrower energy range around the Fermi level, together with the
direct gap and with the density of states (DOS) projected onto atomic
$p_z$ orbitals. The minimum gap of $\sim 1.6$~eV occurs at the S
point, and the electronic states near the band edge are composed
almost entirely of $p_z$ orbitals, with only a small contribution from
other orbitals --~mainly $p_y$~-- at higher energies (not shown).
More precisely, the states at the top of the valence (bottom of the
conduction) band are primarily composed of $p_z$ orbitals on the
C$_{\mathrm{B}}$ (C$_{\mathrm{N}}$) atoms, with smaller but
non-negligible contributions from $p_z$ orbitals on the N and B
atoms. These features guided our choice of the two Wannier basis sets
described in \sect{comp-details}. In particular, the larger set
with one $p_z$ orbital on every atom captures the dominant orbital
character at the band edge as well as the covalent nature of the
system; we will use it in most of our calculations, with the exception
of \sect{direct} where we switch to the minimal basis with $p_z$
orbitals on the carbon atoms only.

\begin{figure}[t]\centering
\includegraphics[width=0.6\linewidth]{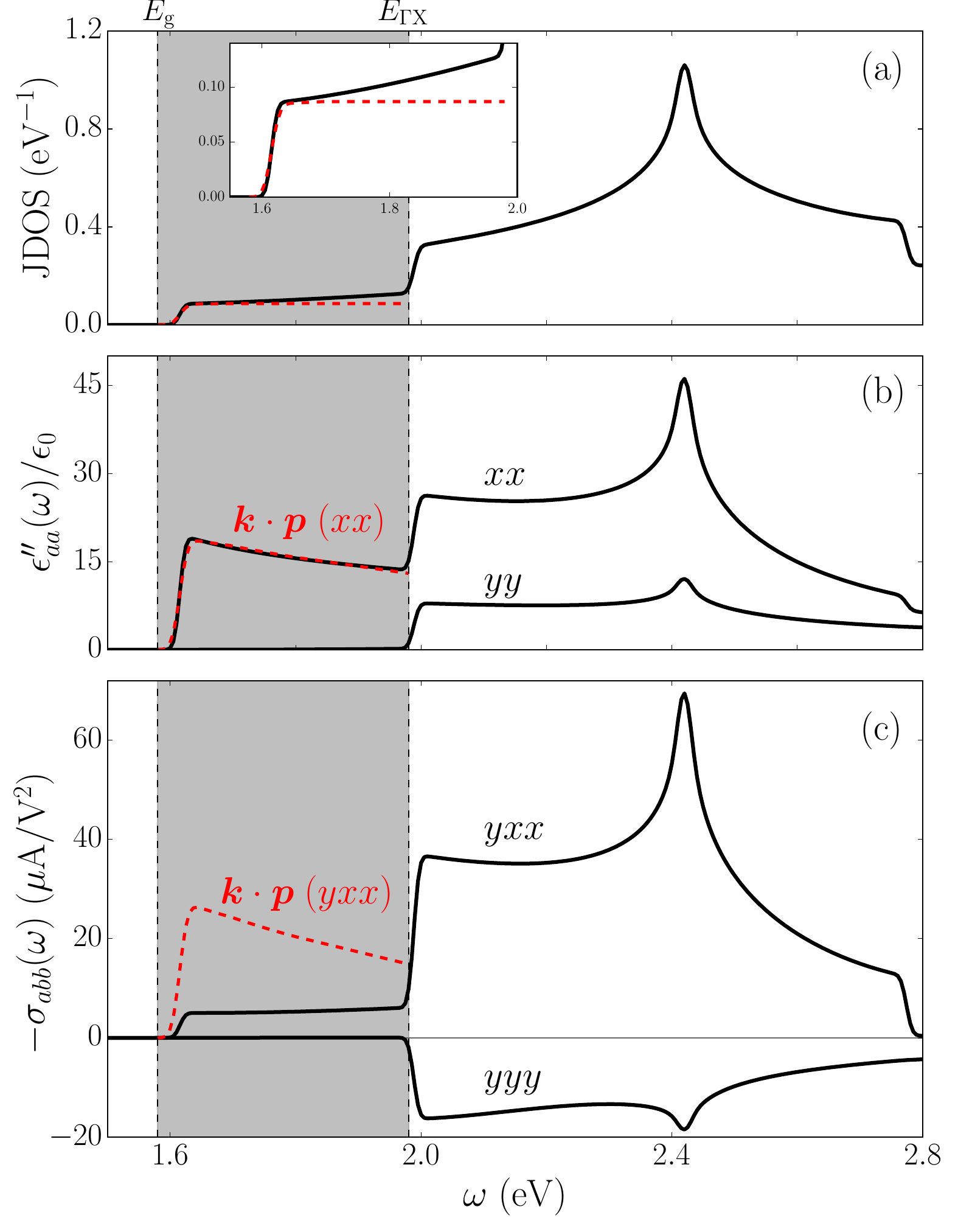}
\caption{(a) Joint density of states (JDOS), (b) imaginary part of the
  dielectric function, and (c) shift-current spectrum.
  Wannier-interpolation results are shown as solid black lines, and
  dashed red lines correspond to the two-band $\kk\cdot\pp$ model
  described in Sec.~\ref{sec:kp}.  The shaded area between
  $E_{\rm g}\sim 1.6$~eV and $E_{\Gamma {\rm X}}\sim 2$~eV is the
  band-edge region, and the inset in (a) shows a blow up of the JDOS
  in that region.}
\label{fig:spectrum-bc2n}
\end{figure}

Next we analyze the optical response tensors evaluated by Wannier
interpolation using our reference basis set. The symmetry-allowed
components of the dielectric function are $\epsilon_{xx}$ and
$\epsilon_{yy}$, and those of the shift photoconductivity
$\sigma_{yxx}$, $\sigma_{yyy}$, and $\sigma_{xxy}=\sigma_{xyx}$; for
the sake of clarity, we focus on $\sigma_{yxx}$ and $\sigma_{yyy}$.
We report bulklike values for $\epsilon^{\prime\prime}_{aa}$ and
  $\sigma_{abb}$, by rescaling the values obtained for the slab by a
  factor of $l/h{\approx 4.0}$ ($l=20$~\AA~is the slab thickness, and
  $h=\sqrt{a_{1}^{2}+a_{2}^{2}}=4.97$~\AA~is the stacking
  distance)~\cite{cook-nc17}.

The
joint density of states (JDOS) is shown in \fig{spectrum-bc2n}(a).
It exhibits van Hove singularities at $E_{\rm g}\sim 1.6$~eV and
$E_{\Gamma {\rm X}}\sim 2$~eV, and a strong peak at $\sim$ 2.4~eV.
These features carry over to the dielectric function in
\fig{spectrum-bc2n}(b).  Between $E_{\rm g}$ and $E_{\Gamma {\rm X}}$,
$\epsilon_{yy}^{\prime\prime}$ is negligibly small; this is caused by
mirror-parity selection rules~\cite{ibanez-azpiroz_directional_2019}
that are exact at $E_{\rm g}$ and hold to a good approximation up to
$E_{\Gamma {\rm X}}$.  Turning to the shift photoconductivity in
\fig{spectrum-bc2n}(c), $\sigma_{yxx}$ also peaks at $\sim 2.4$~eV
while $\sigma_{yyy}$ remains significantly smaller over the entire
range shown.  In the band-edge region, $\sigma_{yyy}\approx 0$ due to
the same selection rules that enforce
$\epsilon_{yy}^{\prime\prime}\approx 0$; in contrast, $\sigma_{yxx}$
is sizeable and shows a step-like feature followed by a plateau.
Overall, the shapes of the {$-\sigma_{yxx}$} and $\sigma_{yyy}$
spectra are reminiscent of those of $\epsilon_{xx}^{\prime\prime}$ and
$\epsilon_{yy}^{\prime\prime}$, respectively.

\subsection{Four-orbital tight-binding model}
\label{sec:4WF}

\begin{figure}[t]\centering
  \includegraphics[width=.99\linewidth]{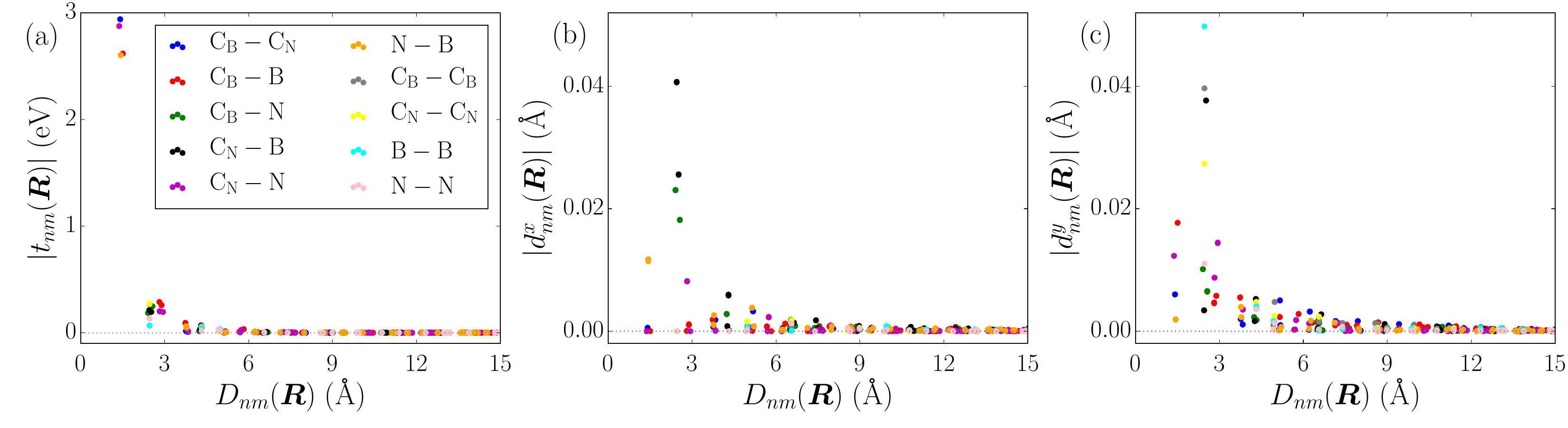}%
  \caption{Decay, as a function of distance between Wannier orbitals,
    of the hopping matrix elements of $\hat H$, $\hat x$, and
    $\hat y$, color-coded by the atomic combination forming the
    orbital pairs. The Wannier basis consists of one $p_z$
      orbital on every atom.}
\label{fig:H_and_r}
\end{figure}

\begin{figure}[!b]\centering
  \includegraphics[width=.95\linewidth]{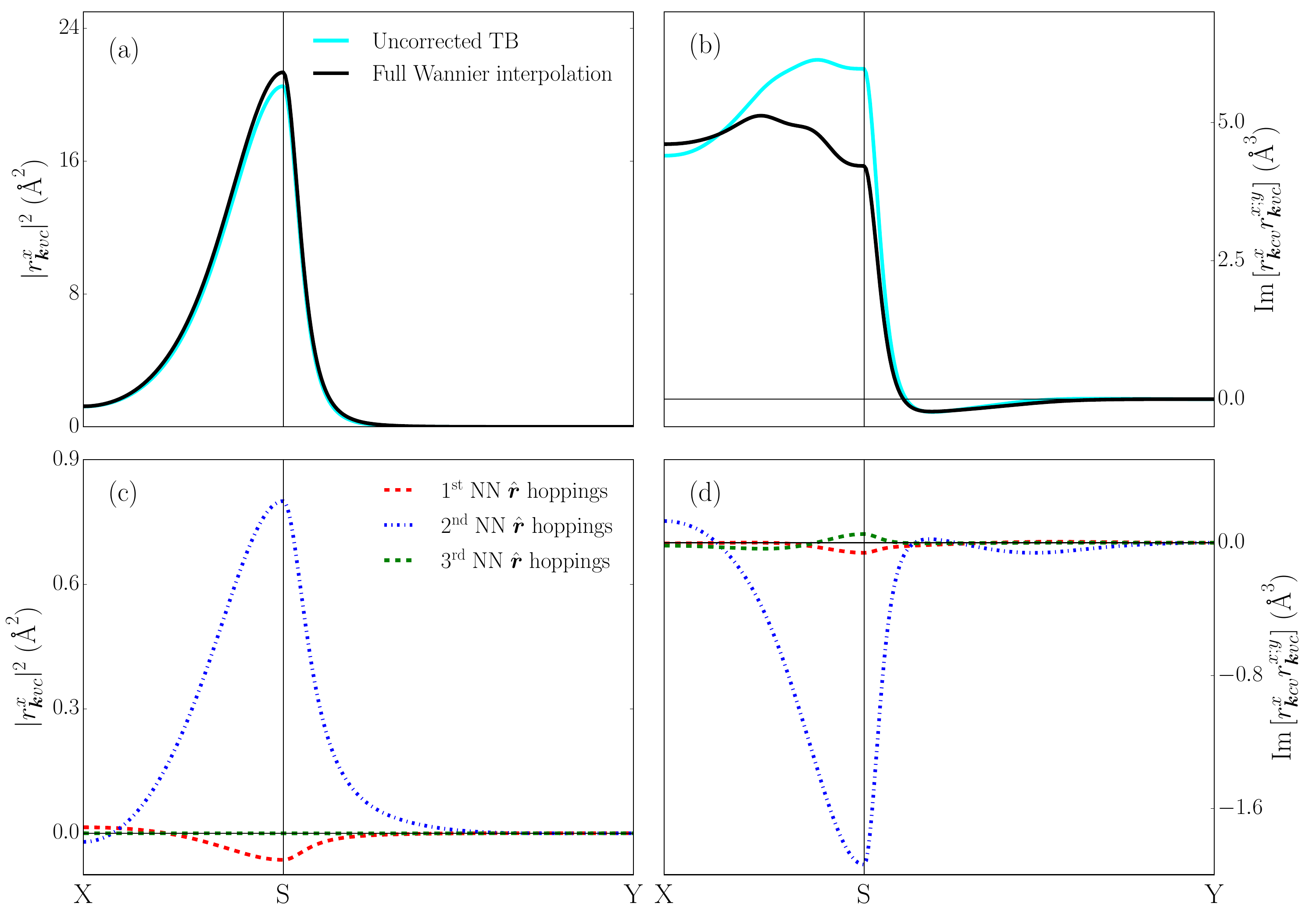}%
  \caption{Dispersion of the $\epsilon^{\prime\prime}_{xx}$ and
      $\sigma_{yxx}$ matrix elements between the top valence band~$v$
    and the bottom conduction band $c$, calculated using a Wannier
    basis with one $p_z$ orbital on every atom.  (a) and (b)
    compare the full Wannier-interpolation matrix elements
      with those obtained within the uncorrected TB approximation. (c)
    and (d) show the contributions of 1st to 3rd
    nearest-neighbor $\hat\rr$-hopping terms to the quantities ploted
    in (a) and~(b).}
\label{fig:kpath4wann}
\end{figure} 

Since our Wannier basis contains one orbital per atom, the $\hat H$
and $\hat\rr$ hoppings in \eqs{e-t}{r-full} are purely
interatomic. Their behavior as a function of distance
\beq
D_{nm}(\RR)=\left|\RR+\btau_m-\btau_n\right|
\eeq
between orbitals is shown in \fig{H_and_r}.  As expected from the
localized nature of WFs, $\hat H$ hoppings decay very rapidly with
distance: the four NN hoppings at $\sim 1.5$ \AA$\;$ dominate over the
rest by more than an order of magnitude, and hoppings beyond 3 \AA$\;$
are negligible.  The behavior of $\hat x$ and $\hat y$ hoppings is
more complex, with 1st NN coefficients being less than half the
size of the dominant 2nd NN ones at $\sim 3$~\AA. The largest
coefficients overall are $\hat y$ hoppings between boron pairs separated along
$x$, but hoppings as distant as $\sim 6$ \AA$\;$ remain sizeable.  The
longer range of $\hat\rr$ hoppings compared to $\hat H$ hoppings,
  as well as their non-monotonic behavior,
seem reasonable given
that the $\hat\rr$ operator grows linearly with distance.

\begin{figure}[t]\centering
  \includegraphics[width=.6\linewidth]{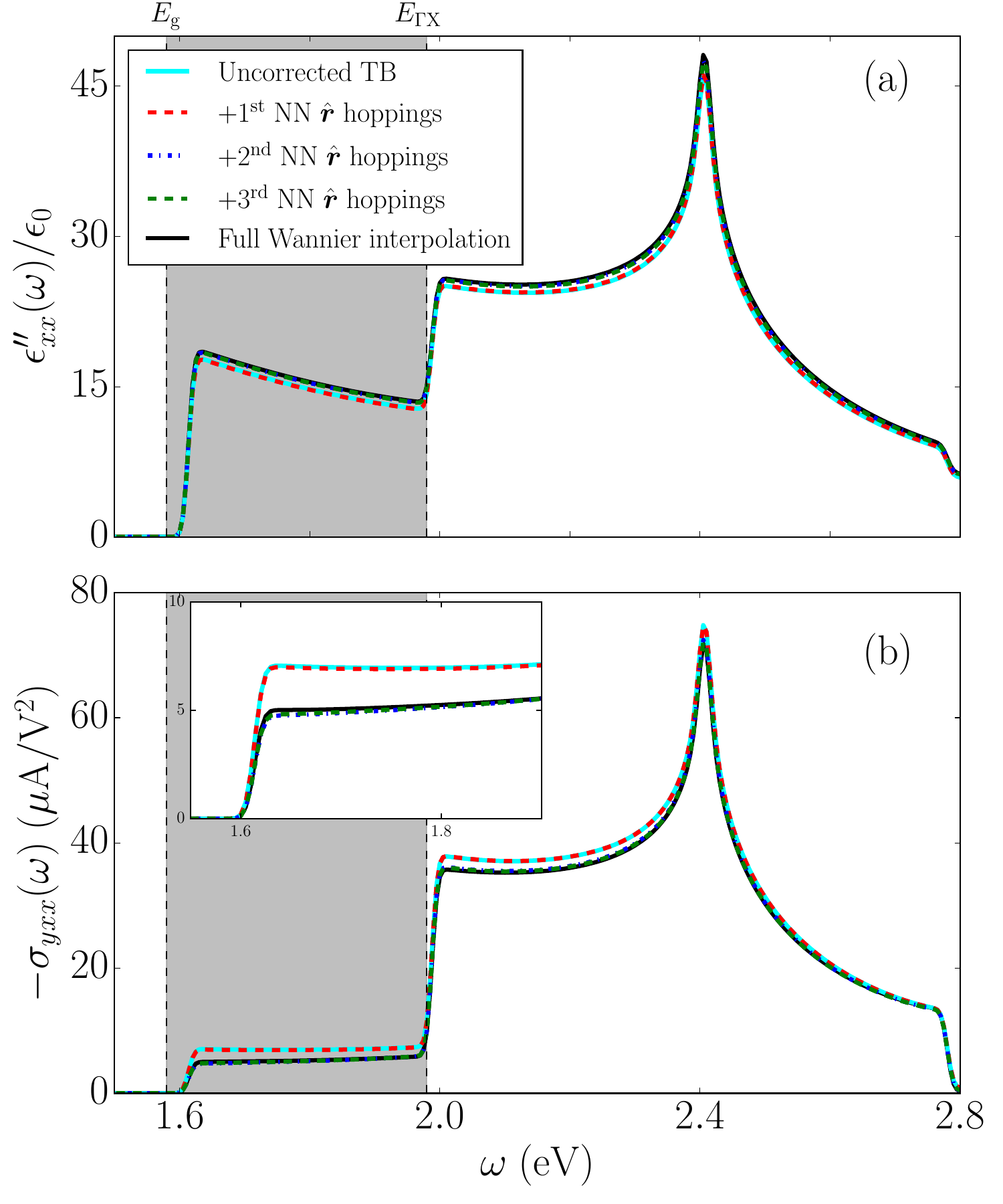}%
  \caption{(a) and (b) $\epsilon_{xx}^{\prime\prime}$ and
    $-\sigma_{yxx}$ spectra, for different levels of truncation of the
    $\hat\rr$ hopping matrix. The Wannier basis and labelling scheme
    are the same as in \fig{kpath4wann}, with the difference that here
    the $\hat\rr$ hoppings are included in a cummulative way. The
    inset zooms in on $-\sigma_{yxx}$ at the band edge. 
         }
 \label{fig:4wann-spectrum}
\end{figure} 

Let us now analyze the impact of $\hat\rr$ hoppings on the calculated
$\epsilon^{\prime\prime}_{xx}$ and $\sigma_{yxx}$ spectra. Before
coming to the full spectra,
we first inspect the associated transition matrix elements between the
top valence and bottom conduction bands.  Their dispersions are
plotted in \fig{kpath4wann}; in panels~(a,b) we compare full
Wannier-interpolation results with uncorrected TB results obtained by
setting all $\hat\rr$ hoppings to zero, and in panels (c,d) we show
the separate corrections to both matrix element from 1st and 2nd NN
$\hat\rr$ hoppings.
The uncorrected TB approximation works very well for the
$\epsilon^{\prime\prime}_{xx}$ matrix element:
the largest relative error, which occurs around
the band edge S, does not exceed 5\%. That approximation is much less
satisfactory for the $\sigma_{yxx}$ matrix element,
especially around S where the relative error reaches $\sim50$\%. In
the case of the $\epsilon^{\prime\prime}_{xx}$ matrix element, the
small corrections to the TB approximation come mostly from 2nd NN
$\hat x$ hoppings;
this is consistent with the behavior of those hoppings in
\fig{H_and_r}(b), where 1st NN terms at $\sim 1.5$~\AA$\;$ are much
smaller than 2nd NN ones at $\sim 3$~\AA. The largest corrections to
the $\sigma_{yyx}$ matrix element again come from 2nd NN $\hat\rr$
hoppings, as expected from Figs.~\ref{fig:H_and_r}(b,c) (the $\hat y$
hoppings contribute through $r_{\kk vc}^{x;y}$).

The full spectra $\epsilon^{\prime\prime}_{xx}$ and $\sigma_{yxx}$,
calculated both with and without $\hat\rr$-hopping corrections, are
displayed in \fig{4wann-spectrum}. Consistent with the preceeding
analysis, those corrections are fairly minor for
$\epsilon^{\prime\prime}_{xx}$, but they are significant for
$\sigma_{yxx}$ in the band-edge region, where $\hat\rr$ hoppings up to
2nd NN have a sizeable impact on the computed spectrum.

\subsection{Two-band models for the band edge}
\label{sec:two-band}

\begin{figure}[t]\centering
  \includegraphics[width=.99\linewidth]{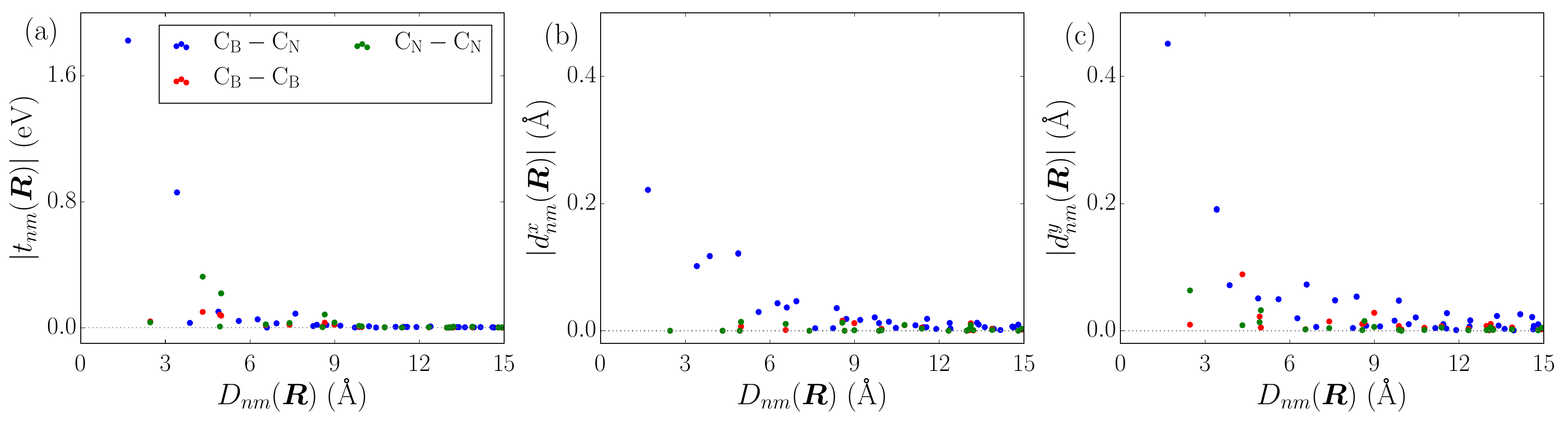}%
  \caption{Same as \fig{H_and_r}, but for a minimal model with $p_z$
    orbitals on the carbon atoms.}
\label{fig:2wann-t_D}
\end{figure}

With the aim of reproducing the electronic and optical properties at
the band edge in the simplest possible way, we now turn to
minimal two-band models.
We consider below two such models constructed in 
  different ways.

\subsubsection{Tight-binding model}
\label{sec:direct}

The first minimal model we consider is a TB model whose basis
  consists of $p_z$-like WFs on the carbon atoms, forming a quasi
one-dimensional chain along $x$ (see \fig{bc2n-crystal-str}).  As
discussed in \sect{comp-details}, this model is expected to yield
acceptable results only at the band edge, where carbon $p_z$ states
are prevalent.

\begin{figure}[!h]\centering
  \includegraphics[width=.95\linewidth]{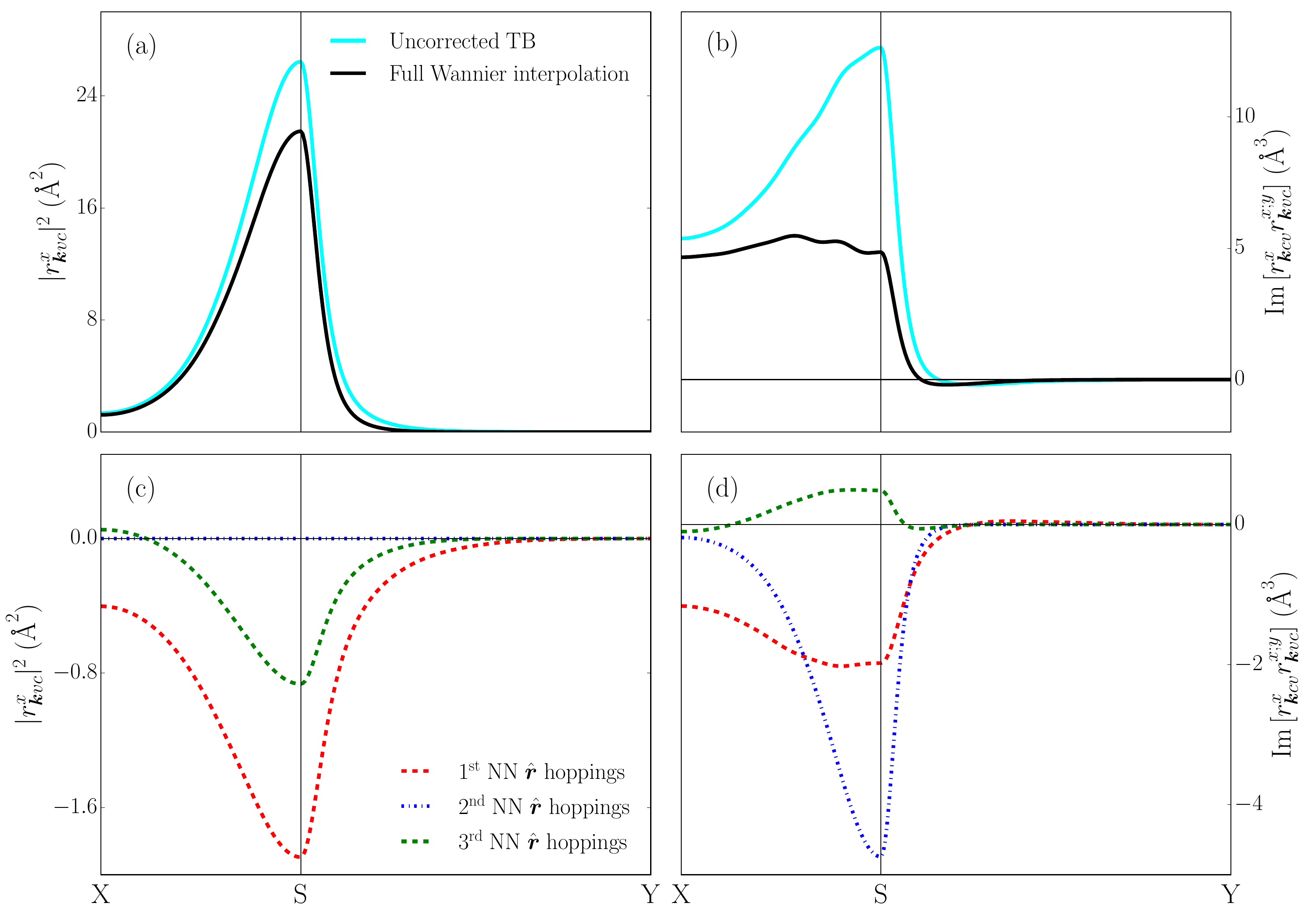}%
  \caption{Same as \fig{kpath4wann}, but for a minimal model with
    $p_z$ orbitals on the carbon atoms.}
\label{fig:kpath2wann}
\end{figure}

 The decay of the $\hat H$ and $\hat\rr$ hoppings in
this model is shown in \fig{2wann-t_D}. Compared with the four-band
model (\fig{H_and_r}), the decay is significantly slower. The largest
$\hat H$ hoppings have similar magnitudes in both models, while the
largest $\hat x$ and $\hat y$ hoppings are an order of magnitude
larger in the minimal model; as a result, $\hat\rr$ hopping
corrections are much more pronounced. This can be seen in the
dispersions of the $\epsilon_{xx}^{\prime\prime}$ and $\sigma_{yxx}$
matrix elements in \fig{kpath2wann}: around S the corrections are
significant already for $\epsilon_{xx}^{\prime\prime}$ [panels~(a,c)],
while in the case of $\sigma_{yxx}$ [panels~(b,d)] they reduce the
matrix element by almost a factor of three, with the largest
corrections coming from 1st and especially 2nd NN $\hat\rr$ hoppings.

\subsubsection{{\boldmath $k\cdot p$} model}
\label{sec:kp}

As an alternative approach for constructing a minimal band-edge model,
we now extract a two-band $\kk\cdot\pp$ effective Hamiltonian from our
reference four-band TB Hamiltonian.  This is motivated in part
by previous works~\cite{cook-nc17,yan_precise_2018,ahn-prx20,xu-npj21},
where two-band $\kk\cdot\pp$ models were used to describe the
band-edge photocurrent response of different types of materials.

Our $\kk\cdot\pp$ model is constructed as described in
Appendix~\ref{appendix}, by expanding the TB
Hamiltonian~\eqref{eq:H-w} to second order in $\kk$ around the S
point, and then applying L\"owdin perturbation theory.  The result is
a transformed $2\times 2$ Hamiltonian $\tilde{H}_\kk$, which we expand
in terms of the identity matrix and of the Pauli matrices as
\beq
\label{eq:Hexp}
\tilde{H}_\kk = \epsilon_{0}(\kk)\mathbb{1}+\sum_{i}f_{i}(\kk)
\sigma_{i}\,.
\eeq
Its band dispersion in \fig{elec-str}
agrees well with the DFT dispersion near the band edge.

Near the band edge, the dielectric function and shift
photoconductivity are given by the product between the transition
matrix elements and the JDOS~~\cite{cook-nc17},
\bea
&\label{eq:epsilon-2band}
\epsilon_{aa}^{\prime\prime}(\omega) = 
\dfrac{\pi e^{2}}{\hbar}
\left| r^a_{\kk vc}\right|^2
N(\omega)\, ,\\
&
\label{eq:sigma-2band}
\sigma_{abb}(\omega) = 
\dfrac{\pi e^{3}}{\hbar^{2}}
\mathrm{Im}\left( r^b_{\kk vc} r^{b;a}_{\kk cv}\right)
N(\omega)\,.
\eea
The $\kk\cdot\pp$ expressions for the matrix elements at a band
  extremum read~\cite{cook-nc17,PhysRevB.87.125425}
\beq
\label{eq:K-twoband}
{\left| r^a_{\kk vc}\right|^2}=
\dfrac{4}{\omega^{2}}\sum_{i}
{\left(f_{i,a}\right)^2}
\eeq
and
\beq
\label{eq:cook-sc}
\mathrm{Im}\left( r^b_{\kk vc} r^{b;a}_{\kk cv}\right)=
-\dfrac{1}{2\omega^{3}}\sum_{ijl}f_{l}f_{i,b}f_{j,ab}\varepsilon_{ijl}\,,
\eeq
where 
the coefficients $f_i$, $f_{i,a} = \partial_a f_{i}$, and
$f_{i,ab} = \partial^2_{ab}f_{i}$ are evaluated at the 
S point using \eq{fi} in Appendix~\ref{appendix}, and
$\varepsilon_{ijm}$ is the Levi-Civita symbol.
Note that these expressions only depend on the $\kk\cdot\pp$
  Hamiltonian, which in turn is constructed from the TB Hamiltonian;
  thus, $\hat\rr$ hoppings are not taken into account when
  evaluating optical matrix elements from a TB-derived $\kk\cdot\pp$
  model.

The $\kk\cdot\pp$ results for the JDOS,
$\epsilon^{\prime\prime}_{xx}$, and $\sigma_{yxx}$ are shown as dashed
red lines in the
shaded regions of \fig{spectrum-bc2n}.  Panel~(a) shows a good
agreement 
with the Wannier-interpolated JDOS around the band gap: the height of
the step-like feature at $E_{\rm g}$ is nicely reproduced, and
although the Wannier-interpolation curve grows monotonically
above $E_{\rm g}$ while the $\kk\cdot\pp$ one stays flat,
the discrepancy is 
small.
In panel~(b), the $\kk\cdot\pp$ curve for
$\epsilon^{\prime\prime}_{xx}$
matches very well the Wannier-interpolation one over the entire
band-edge region: it reproduces not only the step height at
$E_{\rm g}$ but also the subsequent decrease, thanks to the
$1/\omega^{2}$ factor in \eq{K-twoband}.
In contrast,
in panel~(c) the $\kk\cdot\pp$ curve for $\sigma_{yxx}$
deviates considerably from the
Wannier-interpolation one, overshooting it by about a factor of five
at $E_{\rm g}$. This
is in line with our previous finding
  that the $\sigma_{yxx}$ matrix element at the S point gets
  strongly reduced when $\hat\rr$ hopping terms are included.

\section{Discussion}
\label{sec:discussion}

The question of how to evaluate optical matrix elements was debated in
the TB literature until the early 2000s (see
Ref.~\cite{pedersen-prb01} for an overview). It eventually became
clear that the ``minimal TB substitution''
$\hat{\bm v}\rightarrow (1/\hbar) \bnabla_\kk H\W_\kk$ for the
velocity matrix elements leaves out important physics. In particular,
it completely neglects intra-atomic dipole
transitions~\cite{pedersen-prb01}, as well as corrections to
interatomic transitions from off-site dipole matrix elements.
Although the shortcomings of the minimal (or uncorrected) TB approach
to the calculation of optical matrix elements are by now well
understood, that approach continues to be widely used because it
requires no additional parameters beyond the standard ones: on-site
energies, $\hat H$ hoppings, and orbital centers.

The development of Wannier-interpolation schemes by the {\it ab
  initio} electronic-structure community provided an opportunity to
assess the importance of those additional TB parameters (hopping
matrix elements of $\hat\rr$) to various transport and optical
responses in a wide range of materials. In the early works on Wannier
interpolation~\cite{wang-prb06,yates-prb07}, the anomalous Hall
conductivity and magnetic circular dichroism spectrum of bcc Fe were
found to be very well described by the uncorrected TB approach where
all $\hat\rr$ hopping terms are discarded.

More recently, Wannier interpolation has been used to calculate {\it
  nonlinear} optical responses, and more pronounced $\hat\rr$-hopping
corrections were found in some cases, such as the shift
photoconductivities of WS$_2$~\cite{wang-prb17} and especially
GaAs~\cite{azpiroz-prb18}, and the high-harmonic generation spectrum
of monolayer hexagonal BN~\cite{silva-prb19}.  This provided the
motivation for the present work, where we carried out a systematic
study of the impact of hopping matrix elements of $\hat\rr$ on the
linear (dielectric function) and quadratic (shift photoconductivity)
optical responses of monolayer BC$_2$N.

Our findings can be summarized as follows: (i) $\hat\rr$ hoppings
decay more slowly with distance than $\hat H$ hoppings, and 2nd NN
$\hat\rr$ hoppings can be significantly larger than 1st NN ones; (ii)
$\hat\rr$ hoppings are more important for the shift current than for
the dielectric function, indicating that the former is more sensitive
to the spatial structure of the WFs; (iii) the importance of
$\hat\rr$-hopping corrections increases as the number of Wannier basis
orbitals decreases, since the Wannier orbitals tend to be more
extended in minimal bases; (iv) two-band $\kk\cdot\pp$ Hamiltonians
constructed from TB Hamiltonians are likely to
provide a poor description of the
shift-current response at the band edge, due to the neglect of
$\hat\rr$-hopping corrections.

Despite being available
    ``for free'' within the Wannier interpolation framework,
  $\hat\rr$ hoppings are often discarded in 
    Wannier-based calculations of nonlinear optical responses
 (for some recent examples, see
    Refs.~\cite{PhysRevB.100.245206,PhysRevB.97.241118,PhysRevMaterials.4.064602,xu-npj21}).
   Our results indicate that this common
  practice is ill-advised, since the error incurred can sometimes
    be significant.

One question that the present work does not address, and which could
be an interesting direction for future research, is how to devise
reasonable approximations for the hopping matrix elements of $\hat\rr$
in the context of empirical TB theory, where the basis orbitals are
not explicitly available.

\emph{Acknowledgements} --
We thank David Vanderbilt, Michele Modugno and Stepan Tsirkin for discussions.
This work was supported by Grant No.~FIS2016-77188-P
  from the Spanish Ministerio de Econom\'ia y Competitividad.
This project has also received funding from the 
European Union's Horizon 2020 research and innovation
programme under the Marie Sklodowska-Curie grant agreement No 839237
and the European Research Council (ERC) grant agreement No 946629.

\appendix

\section{Construction of the {\boldmath $k\cdot p$} model}
\label{appendix}

{Our $\kk\cdot\pp$ model is obtained by first carrying out a series
  expansion of the four-band TB Hamiltonian
  around the S point, and then using the L\"owdin partitioning scheme
  to reduce it to a two-band Hamiltonian. This is a standard procedure
  for constructing $\kk\cdot\pp$ models starting from TB
  Hamiltonians~\cite{liu-prb13,kaxiras-2015}, and here we describe our
  Wannier-based implementation.}

\subsection{L\"owdin partitioning}
\label{sec:lowdin}

{We begin by reviewing the L\"owdin parititioning
  scheme~\cite{winkler_spin-orbit_2003}.}  Consider a Hamiltonian
\beq
\label{eq:H}
H=H^{0}+H^{\prime}
\eeq
where the eigenvalues $E_{n}$ and eigenfunctions
of $H^{0}$
are known, and $H^{\prime}$ is a perturbation.  Quasi-degenerate
{(L\"owdin)} perturbation theory assumes that the set of
eigenfunctions of $H^0$ can be divided into subsets A and B that are
weakly coupled by $H^{\prime}$, and that we are only interested in
subset A, the ``active'' subspace.  This theory asserts that a
transformed Hamiltonian $\tilde{H}$ exists within subspace A such that
\beq
\label{eq:pert-exp}
\tilde{H}=\tilde{H}^{0}+\tilde{H}^{1}+\tilde{H}^{2} + \cdots,
\eeq
where $\tilde{H}^{j}$ contain matrix elements of $H^{\prime}$ to the
$j$th power. The first three terms are~\cite{winkler_spin-orbit_2003}
\bea
\label{eq:pert-matelem0}
& \tilde{H}^{0}_{mm'} = H^{0}_{mm'},\\
\label{eq:pert-matelem1}
& \tilde{H}^{1}_{mm'} = H^{'}_{mm'},\\
\label{eq:pert-matelem2}
& \tilde{H}^{2}_{mm'} = \dfrac{1}{2}\sum_{l}H^{'}_{ml}H^{'}_{lm'}
\left( 
\dfrac{1}{E_{m}-E_{l}}+\dfrac{1}{E_{m'}-E_{l}}
\right),
\eea
where $m,m'\in \text{A}$ and $l\in \text{B}$.  The approximation
$\tilde{H}\approx \tilde{H}^{0}+\tilde{H}^{1}$ amounts to truncating
$H$ to the A subspace. By
adding $\tilde{H}^{2}$, the
coupling to the B subspace is taken into account.

\subsection{From tight-binding to {\boldmath$k\cdot p$} }
\label{subsec:kp-Wannier}

We shift the origin of $k$ space to a reference 
point (the S point, in our application), and Taylor expand the
Wannier Hamiltonian~\eqref{eq:H-w}
up to second order in $\kk$,
\bea
\label{eq:HW-exp}
H^{(\text{W})}_\kk=H^{(\text{W})}_\OO
+\sum_{a}H_{\OO,a}^{(\text{W})}k_a
+\dfrac{1}{2}\sum_{ab}H_{\OO,ab}^{(\text{W})}k_{a}k_{b} +
\mathcal{O}(k^{3})
\eea
(the notation for $\kk$ derivatives is the same as in
  \eq{cook-sc}).  Applying to
\eq{HW-exp} a similarity transformation $U_\OO$ that
diagonalizes
the $\kk$-independent term, we obtain
the transformed
Hamiltonian
\beq
\label{eq:Hbar}
H_\kk={\overline{H}} + {\sum_{a}\overline{H}_{a}k_{a}
  +\dfrac{1}{2}\sum_{ab}\overline{H}_{ab}k_{a}k_{b}} +
\mathcal{O}(k^{3})\,,
\eeq
where we introduced the notation
$\overline{\mathcal{O}}=U^{\dagger}_\OO\mathcal{O}\W_\OO U_\OO$, and
applied it to $\mathcal{O}_\OO=H_\OO,{H}_{\OO,a},{H}_{\OO,ab}$.  Next
we apply L\"owdin partitioning, choosing the diagonal matrix
$\overline{H}$ as the $H^{0}$ of \eq{H}, and the remaining terms in
\eq{Hbar} as $H^{\prime}$.  Collecting terms in \eq{pert-exp} we get
\beq\label{eq:Htilde}
\begin{split}
\tilde{H}_{\kk mm'} = 
\overline{H}_{mm'} + \sum_{a} \left(\overline{H}_{a}\right)_{mm'}k_{a}
 + \dfrac{1}{2}\sum_{ab}\left[
\left(\overline{H}_{ab}\right)_{mm'} + \left({T}_{ab}\right)_{mm'}
\right]k_{a}k_{b}+ \mathcal{O}(k^{3})\,,
\end{split}
\eeq
where
\beq\label{eq:Tab}
\begin{split}
\left({T}_{ab}\right)_{mm'}=\sum_{l\in B}
\left(\overline{H}_{a}\right)_{ml}\left(\overline{H}_{b}\right)_{lm'} 
\times
\left( 
\dfrac{1}{E_{m}-E_{l}}+\dfrac{1}{E_{m'}-E_{l}}
\right)
= 
\left({T}_{ba}\right)_{m'm}^{*}
\end{split}
\eeq 
and $m,m'\in \text{A}$.  These equations define the effective
$\kk\cdot\pp$ Hamiltonian in the A sector.

Since in our application the A sector contains two bands, we expand
the $\kk\cdot\pp$ Hamiltonian as in \eq{Hexp} in the main text.
To evaluate \eqs{K-twoband}{cook-sc} in the main text, we need the
quantities $f_i$, $f_{i,a}$, and $f_{i,ab}$ at the reference $\kk$
point. Inserting \eq{Htilde} in the expression
$f_{i}(\kk) = (1/2)\text{Tr}\left(\tilde{H}_\kk\cdot
\sigma_{i}\right)$, we find
\begin{subequations}
\label{eq:fi}
\begin{align}
\label{eq:fi0}
f_{i}&=\dfrac{1}{2}
\text{Tr}\left(\overline{H}\cdot \sigma_{i}\right),\\
\label{eq:fia0}
f_{i,a}&=\dfrac{1}{2}\text{Tr}
\left(\overline{H}_{a}\cdot \sigma_{i}\right),\\
\label{eq:fiab0}
f_{i,ab}&=\dfrac{1}{4}\text{Re Tr}
\left[\left(\overline{H}_{ab}+T_{ab}\right)\cdot \sigma_{i}\right]\,,
\end{align}
\end{subequations}
where the traces involve the A-sector blocks of the matrices
  $\overline{H}$, $\overline{H}_a$, and $\overline{H}_{ab}$.

\bibliography{biblio}

\end{document}